\documentclass[12pt]{iopart}
\usepackage{graphicx}

\usepackage{bbm}
\begin{document}

\title{Non-trivial Surface-band Dispersion on Bi(111)}

\author{Yoshiyuki Ohtsubo$^{1}$, Luca Perfetti$^{2}$, Mark Oliver Goerbig$^{3}$, Patrick Le F\`evre$^{1}$, Fran\c{c}ois Bertran$^{1}$ and Amina Taleb-Ibrahimi$^{1,4}$}

\address{$^1$ Synchrotron SOLEIL, Saint-Aubin-BP 48, F-91192 Gif sur Yvette, France}
\address{$^2$ Laboratoire des Solides Irradi\'{e}s, Ecole polytechnique-CEA/DSM-CNRS UMR 7642, F-91128 Palaiseau, France}
\address{$^3$ Laboratoire de Physique des Solides, CNRS UMR 8502, Universit\'{e} Paris-Sud, F-91405 Orsay Cedex, France}
\address{$^4$ URI/CNRS - Synchrotron SOLEIL}
\ead{amina.taleb@synchrotron-soleil.fr}

\date{\today}

\begin{abstract}

We performed angle-resolved photoelectron spectroscopy of the Bi(111) surface to demonstrate that this surface support edge states of non-trivial topology. Along the $\bar{\Gamma}\bar{M}$-direction of the surface Brillouin zone, a surface-state band disperses from the projected bulk valence bands at $\bar{\Gamma}$ to the conduction bands at $\bar{M}$ continuously, indicating the non-trivial topological order of three-dimensional Bi bands.
We ascribe this finding to the absence of band inversion at the $L$ point of the bulk Bi Brillouin zone. According to our analysis, a modification of tight-binding parameters can account for the non-trivial band structure of Bi without any other significant change on other physical properties.
\end{abstract}

\submitto{\NJP}

\section{Introduction}
\label{sec:1}

Ever since the experimental realization of a two-dimensional (2D) topological insulator (TI) in 2007 \cite{Konig07}, topological order (TO) has attracted 
a tremendous interest in the band-structure characterization of condensed-matter systems \cite{Fu07, Kane10, Zhang12}. After these 2D materials, TO
has also been shown to be relevant in the description of three-dimensional (3D) systems, such as Bi$_x$Sb$_{1-x}$, which is the first material 
experimentally detected as a non-trivial TI in 3D \cite{Hsieh08, Hsieh09, Nakamura11}. The probably most salient common feature of TIs in 2D and 3D
consists of the current-carrying edge or surface states (SS), respectively, that are due to a position-dependent change of a characteristic 
topological invariant. This yields the local vanishing of the band gap and therefore precisely the current-carrying states located in the regions
where the invariant changes and that are just the surfaces or interfaces with different materials.
As long as time-reversal symmetry (TRS) is preserved in these systems, the edge and surface states are topologically protected
from a band-gap opening and reveal, in 2D, the quantum spin Hall effect, whereas in 3D they are characterized by an unconventional spin texture 
\cite{Hsieh09} and prohibited electronic backscattering \cite{Roushan09}.

A large class of 3D TIs consists of Bi-based compounds. 
Especially Bi$_x$Sb$_{1-x}$ has almost the same 3D bands as pure Bi and Bi is therefore one of the most frequent examples invoked in the theoretical discussion of TO in 3D electronic structures.
Nevertheless, the most popular models \cite{Golin68_1, Liu95} indicate that pure Bi possesses trivial TO \cite{Fu07, Kane10, Teo08} and various first-principles calculations \cite{Koroteev04, Hirahara06, Koroteev08, Zhang09} show a SS dispersion consistent with this absence of TO.
According to these works, no spin-split SS connects the projected bulk valence bands (BVB) and bulk conduction bands (BCB) between surface time-reversal-invariant momenta (TRIM) on surface Brillouin zone (SBZ).
Thus, pure Bi has been regarded as a typical semimetal with trivial TO.
Despite this common understanding, the TO of Bi is still an open question. Indeed, the dispersion of SS calculated at one surface TRIM on Bi(111) (namely the $\bar{M}$ point) is in qualitative disagreement with the one observed by angle-resolved photoelectron spectroscopy (ARPES) experiments, both in 3D bulk samples \cite{Ast03} and in thin films \cite{Hirahara06, Hirahara08, Takayama11}: 
a SS was observed to be connected to BCB in contrast to the above-mentioned theoretical calculations. 
The interference between the top and bottom surfaces could explain the discrepancy between theory and experiment at $\bar{M}$ for the thin-film case.
However, a prominent coupling between opposite surfaces is not a reasonable assumption for such a discrepancy in the explanation in bulk crystals.
The latter should rather be attributed to the DFT calculations that may not be accurate enough to reproduce the band dispersion in the low-energy regime and thus to identify TO either in the bulk Bi crystal or at the SS. 
Curiously, the finite difference between the empirical electronic states \cite{Brown63, Tichovolsky69, Hiruma83} and DFT results \cite{Hirahara07} has never been discussed by the previous literature.

In this work, we revisit the SS bands on Bi(111) by synchrotron-radiation ARPES with various photon energies and incident photon polarizations.
Our result demonstrates the non-trivial topology of the SS bands on Bi(111) around $\bar{M}$: a SS branch is connected to both BVB and BCB continuously along the line connecting the SBZ points $\bar{\Gamma}$ and $\bar{M}$.
This discrepancy between the theory and experiment can be solved by assuming the non-trivial TO of the bulk bands of Bi. We propose a modification of the common tight-bonding (TB) model \cite{Liu95}, which can explain the non-triviality of bulk Bi without contradiction against previous experimental results.
The remaining parts of the paper are organized as follows. In section \ref{sec:2}, we introduce the experimental methods, and we present our results
in section \ref{sec:3}. Section \ref{sec:4} is devoted to detailed discussions, both from the point of view of a possible topological classification
of semimetals (\ref{sec:4.1}) and in view of the modification of the TB parameters to account for our findings (\ref{sec:4.2}). We present our
conclusions in section \ref{sec:5}.

\section{Experimental methods}
\label{sec:2}

ARPES measurements were performed at the CASSIOPEE beamline (SOLEIL, France) with a hemispherical photoelectron spectrometer (VG Scienta R4000) with an energy resolution of 7 meV and an angular resolution of 0.1$^\circ$.
Linearly and circularly polarized lights with photon energies from 12 to 35 eV were used.
The spectra were measured at 7.5 K in a base pressure below 1 $\times$10$^{-10}$ mbar.
The (111) surface of Bi single crystal has been obtained by Ar ion sputtering and annealing cycles of a single crystal until a sharp (1$\times$1) low-energy electron diffraction (LEED) pattern was observed. 
The cleanliness of the surfaces was also checked by Auger electron spectroscopy.

\section{Results}
\label{sec:3}

Figure 1 (a) shows the Fermi contour on Bi(111).
The hexagonal electron pocket ($\Sigma_2$) and petal-like hole pockets ($\Sigma_1$) are clearly observed.
They are almost identical to those previously observed on Bi(111) \cite{Hirahara06, Takayama11, Ast01}. 
Figure 1 (b) shows the SS band dispersion along $\bar{\Gamma}\bar{M}$.
Both $\Sigma_1$ and $\Sigma_2$ loose their intensity near $\bar{\Gamma}$, indicating that they merge into projected BVB there.
Continuous dispersions of $\Sigma_1$ and $\Sigma_2$ without crossing with other SSs above the Fermi level ($E_{\rm F}$) are theoretically expected \cite{Teo08, Koroteev04} and have recently been observed by a time-resolved ARPES experiment \cite{Ohtsubo12}.

Figure 2 shows the observed SS dispersion around $\bar{M}$.
In Fig. 2(a), the Fermi contour around $\bar{M}$ is shown. 
Figure 2(c) is the SS band dispersion near $E_{\rm F}$ measured along $\bar{\Gamma}\bar{M}$.
Both SS bands didn't show any energy shift with incident photon energy, indicating their two-dimensional nature.
The lower branch of SS connects to the hole pocket $\Sigma_1$ observed around $\bar{\Gamma}$ (Fig. 1(b)).
$\Sigma_1$ disperses from 60 meV to 150 meV and merges into the projected bulk bands around 0.7 \AA $^{-1}$.
The upper branch ($\Sigma_2$) appears below $E_{\rm F}$ at 0.55 \AA $^{-1}$ and looses its intensity in proximity of $\bar{M}$.
As clearly shown in the 2nd-derivative plot in Fig. 2(d), they do never cross each other at $\bar{M}$.
Both $\Sigma_1$ and $\Sigma_2$ are not clearly visible at $\bar{M}$, suggesting that they are degenerated with bulk bands nearby $\bar{M}$.

Figure 2 (e, f) shows the ARPES image taken along $\bar{M}\bar{K}$.
The broad emission feature near $E_{\rm F}$ shows no dependence on photon energy.
Since there are no SS below $E_{\rm F}$, this photoelectron signal should originate from the projected bulk bands.
These bulk bands near $E_{\rm F}$ strongly depend on the polarization of the incident photon.
It could explain why they were not observed in a previous work \cite{Ast04}.
We overlapped the calculated projected bulk-band edges (dashed lines), based on the TB parameters of Ref. \cite{Liu95}, onto the ARPES plots.
In Fig. 2 (e, f), the edge of the calculated bands show a good agreement with the edge of ARPES intensity.
In addition, the calculated position of the edges agrees with the area where $\Sigma_1$ and $\Sigma_2$ become diffuse in Fig. 1 (b) and Fig. 2 (c).
These agreements show that there are almost negligible band bending near the Bi(111) surface.
Based on these experimental and calculated results, we conclude that the SS branch $\Sigma_2$ is connected to BVB and BCB at $\bar{\Gamma}$ and $\bar{M}$, respectively, and disperses continuously between the two points.
On the other hand, $\Sigma_1$ merges into BVB both at $\bar{\Gamma}$ and $\bar{M}$.

\section{Discussion}
\label{sec:4}

In order to discuss TO of semimetallic Bi, we first classify TO of semimetals both from their 3D band structure and SS dispersions.
Then, we apply this reasoning to Bi(111) and discuss the detailed 3D band structure of Bi based on the TB model \cite{Liu95} with modified hopping parameters.

\subsection{Topological classification of semimetals}
\label{sec:4.1}

Notice that the topological classifications we adopt here for semimetals have originally been proposed for insulators, in which the Fermi level
is always situated within a bulk gap separating the BVB from the BCB at any wave vector in the first Brillouin zone (BZ). In the present section, we 
justify this use in the framework of semimetals with overlapping energy bands that are though separated by an energy gap for every wave vector in the 
first BZ. We thus consider energy bands that do not cross in spite of their energy overlap that leads to electron and hole pockets, as in 
the present case of Bi. In such a case, one may continuously connect the semimetal to a true insulator by a transformation
\begin{equation}\label{eq:PFS}
H({\bf k}) \rightarrow H'({\bf k}) = H({\bf k}) + \epsilon ({\bf k}) \mathbbm{1},
\end{equation}
where $H({\bf k})$ is the reciprocal-space Hamiltonian in matrix form that yields the energy bands of Bi, e.g. within a TB description. The energy
function $\epsilon ({\bf k})$ simply shifts all energy bands as a function of the wave vector ${\bf k}$ in such a manner that one eventually obtains 
a bulk insulator, and $\mathbbm{1}$ is the one matrix.
Notice that, in order to maintain TRS, one needs to have 
$\epsilon ({\bf k})=\epsilon (-{\bf k})$. One may identify $-\epsilon({\bf k})$ with a pseudo Fermi surface, that is an energy surface 
in the original semimetal situated in between the adjacent energy bands that cross the true Fermi level \cite{Zhang12, Footnote1}.
 The most crucial point to realize is that the term $\epsilon ({\bf k}) \mathbbm{1}$ in Eq. (\ref{eq:PFS})
does not alter the eigenstates associated with the different bands and therefore does not affect the associated topological properties of the bands.
Therefore, the Chern numbers of the different bands, as well as the $Z_2$ invariants calculated at the TRIMs, 
remain unchanged with any $\epsilon ({\bf k})$.
One may use the transformation
(\ref{eq:PFS}) in a topological classification of semimetals in the same manner as for insulators, as long as one refers to the pseudo Fermi surface
instead of the true Fermi level. 
We emphasize that the pseudo Fermi surface is thus a purely theoretical construct that simplifies the above-mentioned classification of semimetals, whereas it coincides with the physical Fermi surface in the case of insulators.

In order to illustrate the above arguments with respect to the SS behavior, we consider four typical cases depicted in Figs. 3(a-d), corresponding to a 
TI (a), a non-trivial semimetal (b), a trivial band insulator (c), and a semimetal (d) that is trivial from a topological point of view. In the case of a 
TI with TRS, there are necessarily SS (solid lines) that connect the projected BVB and BCB and that cross the Fermi level (dashed line)
an odd number of times. The 
case depicted in Fig. 3(a) shows three Fermi-level crossings, but one may obtain a single crossing if one continuously lowers the lower SS. The associated
$Z_2$ invariant is simply interpreted as the parity of the Fermi-level crossings, and it is conserved for all TRS perturbations that alter the SS. This
TI is furthermore continuously connected, via the transformation (\ref{eq:PFS}), to the semimetallic case sketched in Fig. 3(b), where a part of the BCB
shifts below the Fermi level in the vicinity of the TRIM $\Lambda_a$, thus forming an electron pocket, whereas a part of the BVB floats up around 
$\Lambda_b$ such as to form a hole pocket. As already mentioned, the transformation (\ref{eq:PFS}) does not alter the topological characterization of
the different bands because of the trivial role played by the term $\epsilon ({\bf k}) \mathbbm{1}$, such that the same topological classification
may be adopted to the semimetal as for the original TI. However, the topological invariants must now be defined with respect to the pseudo Fermi surface,
which is depicted by the thick dotted line in Fig. 3(b). Again the number of crossings between the SS and the pseudo Fermi surface is an odd integer 
related to the $Z_2$ invariant, and there must be a continuous SS between the BVB and the BCB. Notice, however, that
this invariant does not give insight into the number of SS crossings with the true Fermi level (dashed line). Indeed, one may continuously deform the SS in such a manner as to position them below the true Fermi level.

Similarly to the above case, one may classify a topologically trivial semimetal as one that is continuously connected, via the transformation 
(\ref{eq:PFS}), to a trivial band insulator depicted in Fig. 3(c). In the presented case, there are four SS crossings with the Fermi level, but a 
continuous deformation or a TRS perturbation may result in two or zero Fermi-level crossings. Again this even parity is preserved for a trivial semimetal
when defined with respect to a pseudo Fermi surface [not depicted in Fig. 3(d)], but again there is no information about the number of crossings with the 
true Fermi level. We finally emphasize that, as in the insulator cases (a) and (c), there is no continuous TRS connection between the topological (b) and
the trivial semimetal (d) unless one allows for band crossings between the BVB and the BCB. This would correspond to a gap closing in the insulator case, 
accompanied by a band inversion at the band crossing. 

Notice that there is a possible ambiguity of the SS counting with the closing of projected bulk band gap on SBZ, in contrast to the bulk BZ: $e$.$g$. around $\bar{X}_2$ on Sb(110) \cite{Bianchi12}.
Fortunately, this is not the case for Bi(111), since there is a finite size of projected bulk band gap at any in-plane wave vector ($k_{\parallel}$) on SBZ.
Based on these methods, the observed SS dispersion on Bi(111), that connects the projected BVB and BCB continuously, clearly indicates the non-trivial TO of bulk bands of Bi as the equivalent case to Fig. 3(b).

It should finally be stressed that the non-trivial TO of semimetal, contrary to that in TIs, does not provide insight into the transport properties of the material, for two major reasons.
First, as mentioned above, it does not guarantee the existence of metallic SS because on a surface of a non-trivial semimetal, SS can be connected to both BVB and BCB without crossing $E_{\rm F}$.
Second, the transport properties of semimetals would be dominated by the bulk bands, which provide an electronic density of states that outcasts that of the SS because of its larger spatial dimension.
However, from the point of views of the mere SS band dispersion, TO of semimetal is equivalent to those for insulators.
Furthermore, even if the topologically protected SS in the present semimetallic system is not responsible for transport measurements, it is noteworth to mention that the SS persists in the case of a continuous deformation, e.g. in the presence of pressure or doping. If no band-contact points are generated and if the electron and hole pockets are suppressed by the deformation, the system may eventually evolve into a true insulating state of the same topological properties, in which case the SS becomes a true metallic surface that would be responsible for electronic transport.

\subsection{Tight-binding description of Bi(111) bands}
\label{sec:4.2}

In order to understand TO of Bi and the dispersion of SS on Bi(111) without ambiguity, we assumed a non-trivial band structure of bulk Bi by slight modifications on TB parameters in Ref. \cite{Liu95}.
TO of bulk Bi is predicted to be trivial since the band inversion occurs at every bulk TRIM \cite{Fu07, Kane10, Teo08}.
However, there is an ambiguity on a band inversion at $L$ in bulk Bi \cite{Golin68_1}, because of the small size of the gap (15 meV) and almost symmetric dispersion of BVB and BCB in the vicinity of this point \cite{Brown63}.
Figure 4 (a) depicts the band evolution at $L$ when varying one TB parameter $V'_{ss\sigma}$, which represents the second-nearest-neighbor interaction between 6$s$ orbitals.
It is shown that the band inversion at $L$ is absent for the values of $V'_{ss\sigma}$ larger than -0.21 eV, and hence Bi has non-trivial TO in this region.
With a slightly larger value of $V'_{ss\sigma}$ (-0.015 eV), the same size of band gap is obtained, as indicated by an arrow in Fig. 4 (a).
Figure 4 (b) shows the bulk valence bands around $E_{\rm F}$ based on TB parameters in Ref. \cite{Liu95} and this work (the other parameters were unchanged).
The obtained bulk bands around $E_{\rm F}$ show almost no change.
Although there are finite difference between experiments and our calculation, such as an underestimation of a few meV in the electron-pocket size at $L$, it is possible to construct the TB model with non-trivial TO which is exactly consistent to experiments, such as pocket size and pocket shape, by modifying two or more parameters.

The non-trivial TO of Bi is not in conflict with previous experimental results. While magnetoreflection measurements \cite{Brown63, Hiruma83} can determine the shape and size of Fermi surface, it does not provide a direct information about band inversion at $L$. 
Only two indirect suggestions were published so far: one from the sign of the gap from the magnetoreflection compared with Bi$_{1-x}$Sb$_x$ \cite{Tichovolsky69} and the other from electron-hole recombination time measurement \cite{Lopez68}.
However, none of them measured the parity of the bulk bands at $L$ directly.

Disagreements of SS dispersion on Bi(111) between the theoretical prediction based on DFT and experimental results could be also explained by the ambiguity of the bulk band inversion at $L$.
DFT calculation shows a qualitative agreement of bulk band dispersion with experimental results, such as the existence of hole pocket at $T$ and electron pocket at $L$, but there are finite size differences between them \cite{Hirahara07}. 
Although they are only few tens of meV, it is large enough to make a spurious TO estimation of bulk Bi, as discussed above.
The degeneracy between $\Sigma_1$ and $\Sigma_2$ at $\bar{M}$ \cite{Hirahara06, Koroteev08} would naturally outcome based on the trivial TO.

Interestingly, calculations done in thin films of Bi better reproduce the non-trivial topology of SS.
This observation suggests that the interference between the top and bottom surfaces can re-inverse the inverted bulk bands, as observed at the HgTe quantum well \cite{Konig07, Chu11}.
Such re-inversion of bulk bands would occur at $L$, because the size of the band gap is smallest there when compared to the other bulk TRIMs.
However, the systems corresponding to the bulk crystal, such as an asymmetric slab \cite{Hirahara06, Koroteev08} and a semi-infinite crystal \cite{Zhang09}, do not include such interference effect, and hence the degeneracy of SS at $\bar{M}$ has been predicted so far. 

Alternatively, one can expect that the surface relaxation on Bi(111) \cite{Monig05} is the cause of the absence of the bulk-band inversion at $L$.
A DFT calculation showed that TO of surface layers and deeper bulk layers could be different \cite{Liu11}.
Moreover, the possibility of topological phase transition on a Bi(111) thin film driven from the structural distortion is claimed very recently \cite{Hirahara12}.
In order to examine this possibility, we applied structural distortions on the TB model obtained above along both in-plane and out-of-plane directions.
Figures 4(d) and (e) show bulk band evolutions at $L$ with structural distortions based on the TB parameters which make trivial and non-trivial TO, respectively.
In both cases, in-plane distortions can invert the bulk bands and hence cause the topological phase transition.
This result agrees with ref. \cite{Hirahara12}.
In contrast, out-of-plane distortions cause no band inversion.
Since there is only an out-of-plane distortion in our case of single-crystal Bi, the surface relaxation of interlayer distances cannot explain the non-trivial TO of Bi (111).
Note that our TB modification, which makes Bi non trivial, does not conflict against the results in ref. \cite{Hirahara12}, since our TB parameters with in-plane strain also predict the non-trivial TO.
The only difference is that Bi is non-trivial even without any strain. 
These results suggest that the non-trivial TO observed on Bi(111) is not due to the atomic structure of (111) surface but the intrinsic character of Bi bulk bands.

The SS dispersions on other Bi surfaces measured by ARPES \cite{Agergaard01, Hofmann05, Wells09} are not in conflict with the non-triviality of Bi.
The spin-polarized non-trivial state was observed on (114) \cite{Wells09} and SS observed on (100) between $\bar{\Gamma}$ and $\bar{M}$ could also be interpreted as non-trivial states: a pair of of SS is in BVB at $\bar{\Gamma}$, but only one branch of them is observed to merge into BVB at $\bar{M}$, at least below $E_{\rm F}$ \cite{Hofmann05}.
While the SS on Bi(110) is also observed by ARPES \cite{Agergaard01}, it is impossible to consider the band inversion at $L$ from (110), since there are no projected bulk band gap at $\bar{X}_2$, in the same manner as for Sb(110) \cite{Bianchi12}.

\section{Conclusions}
\label{sec:5}

In conclusion, we show a SS band which is connected to both BVB and BCB continuously between surface TRIMs on Bi(111), indicating the non-trivial TO of the Bi bulk bands. In contrast to the common understanding, only non-inverted bulk bands at $L$ can explain our results without any contradiction.
We propose a modified TB model which is consistent to the previous experimental results and which account for the non-trivial bulk band structure of Bi.

\ack{The authors gratefully acknowledge stimulating discussions with Gilles Montambaux, Jean-No\"el Fuchs, and Fr\'ed\'eric Pi\'echon from the 
theory group at LPS, Orsay.}

\newpage

\begin{figure}
\begin{center}
\includegraphics[width=80mm]{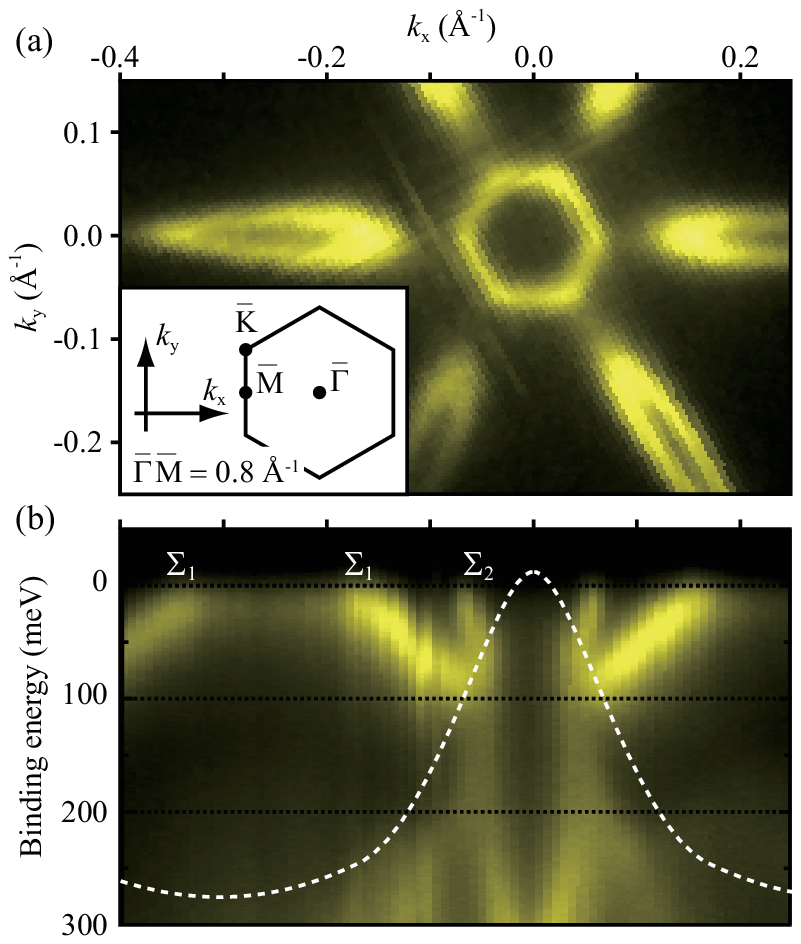}
\end{center}
\caption{\label{fig1} 
(a) Angle-resolved photoelectron spectroscopy (ARPES) intensity plot at the Fermi level ($E_{\rm F}$) measured with $h\nu$ = 15 eV at 7.5 K. The inset shows the surface Brillouin zone and our definition of $k_x$ and $k_y$ directions.
(b) The ARPES intensity plot taken near $\bar{\Gamma}$ along $\bar{\Gamma}\bar{M}$. Dashed line shows the upper edge of the projected bulk bands (see text).
Binding energy is defined relative to $E_{\rm F}$.
}
\end{figure}

\begin{figure}
\begin{center}
\includegraphics[width=80mm]{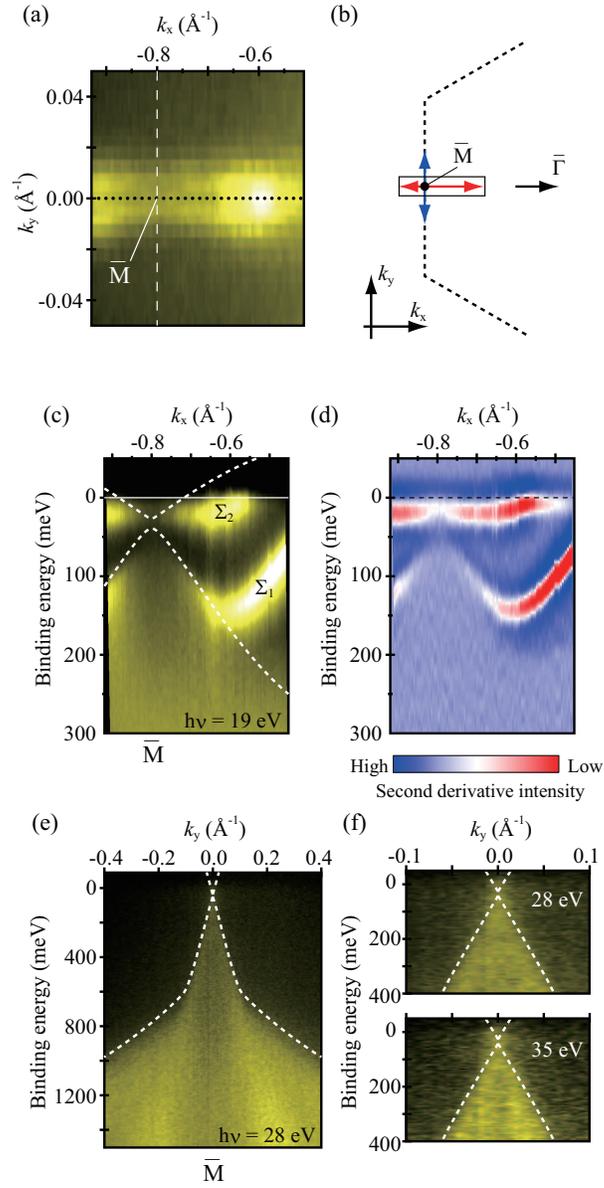}
\end{center}
\caption{\label{fig2} 
(a) ARPES intensity plot at $E_{\rm F}$ measured with $h\nu$ = 19 eV.
(b) Schematic drawing showing the scanned region in SBZ. Solid square and horizontal/vertical arrows represents the regions shown in (a), (c, d), and (e, f), respectively.
(c) ARPES image along $\bar{\Gamma}\bar{M}$. Dashed lines represent the lower and upper edges of the projected bulk bands.
(d) Second-derivative ARPES image at the same region as (c).
(e) ARPES image along $\bar{M}\bar{K}$. Dashed lines are the edges of the projected bulk bands.
(f) ARPES image along $\bar{M}\bar{K}$ in the proximity of $\bar{M}$. 
}
\end{figure}

\begin{figure}
\begin{center}
\includegraphics[width=80mm]{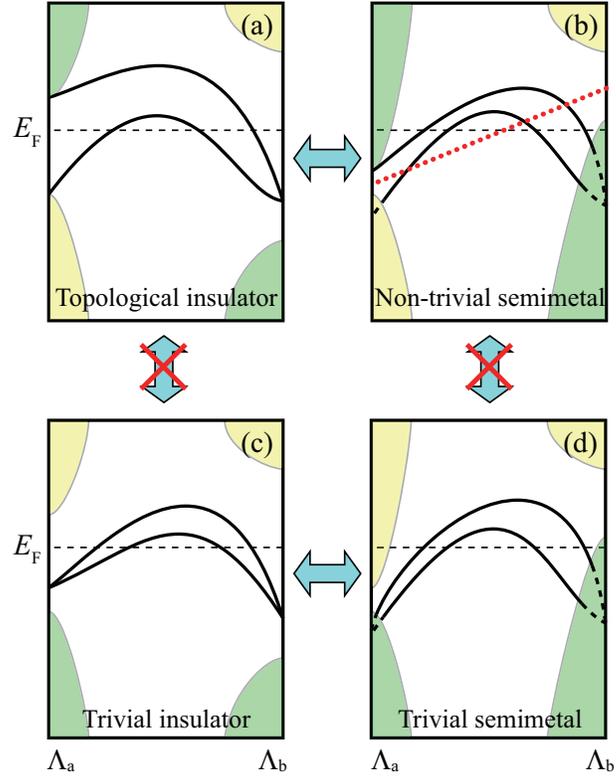}
\end{center}
\caption{\label{fig3} 
(a-d) Schematic representations of SS band dispersions between surface TRIMs $\Lambda_a$ and $\Lambda_b$ on the surface of (a) typical TI
(b) non-trivial semimetal (as Bi(111) observed experimentally) (c) trivial semimetal, and (d) trivial insulator. 
The shaded area represents the projected bulk bands. Colors of them represent parity eigenvalues for each bulk bands at corresponding bulk TRIMs: yellow for -1 and green for +1.
A perturbation can deform a bulk band structure from (a) to (b) or (c) to (d) without any bulk-band inversion, but can neither from (a) to (c) nor (b) to (d).
For a detailed discussion, see text.
}
\end{figure}

\begin{figure}
\begin{center}
\includegraphics[width=80mm]{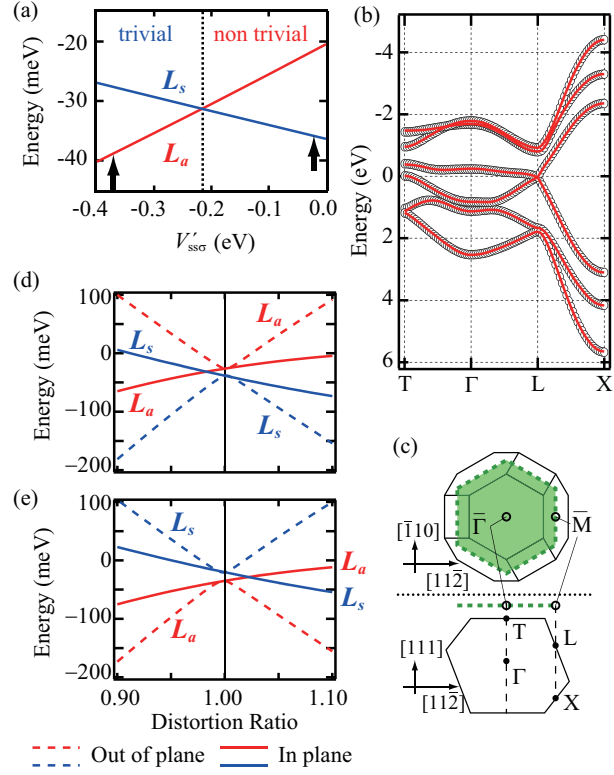}
\end{center}
\caption{\label{fig4} 
(a) Band evolution at $L$ based on tight-binding (TB) model. The origin of energy axis corresponds to $E_{\rm F}$. Arrows indicate the TB parameter in Ref. \cite{Liu95} and the value which provides the same size of band gap with the opposite TO.
(b) Band structure of Bi along some symmetry lines. Lines (trivial) and circles (non-trivial) are calculated based on the TB parameter indicated by arrows in (a). The other parameters are the same as ref. \cite{Liu95}.
(c) Schematic drawing of three-dimensional Brillouin zone (solid line) of the Bi single crystal and its projection onto the (111) surface Brillouin zone (dashed line).
(d) Band evolution at $L$ with in-plane (solid lines) and out-of-plane (dashed lines) lattice distortion based on TB parameters in Ref. \cite{Liu95}.
(e) Same as (d) but TB parameters which result in non-trivial bulk bands without distortion.
}
\end{figure}


\begin{thebibliography}{99}
\bibitem{Footnote1}
In Ref. [3], the term ``imaginary Fermi surface'' has been used.
\end{thebibliography}
\end{document}